\documentclass[useAMS,usenatbib]{mn2e}
\pdfoutput=1
\usepackage[T1]{fontenc}
\usepackage{amssymb}

\usepackage[pdftex]{graphicx}

\title[High-resolution models of solar granulation]{High-resolution models of solar granulation: the 2D case}
\author[H. Muthsam et al.]{H.J. Muthsam$^{1}$\thanks{E-mail:
herbert.muthsam@univie.ac.at },  B. Löw-Baselli$^{1}$, Chr. Obertscheider$^{1}$, M. Langer$^{1}$,
P. Lenz$^{{1},{2}}$, F. Kupka$^{3}$ \\
$^{1}$Faculty of Mathematics, University of Vienna, Nordbergstrasse 15, A-1090 Vienna, Austria\\
$^{2}$Institute for Astronomy, University of Vienna, Türkenschanzstrasse 17, A-1180 Vienna, Austria\\
$^{3}$Max Planck Institute for Astrophysics, Karl Schwarzschild Strasse 1, D-85741, Garching, Germany }

\begin{document} 

\date{???}

\pagerange{\pageref{firstpage}--\pageref{lastpage}} \pubyear{?}

\maketitle

\label{firstpage}

\begin{abstract}
Using advanced numerical schemes and grid-refinement we present 2D high-resolution models of solar granulation with particular emphasis to downflowing plumes. In the high-resolution portion of our simulation, a box measuring   $1.97 \times 2.58 \mbox{ Mm}^2$ (vertical $\times$ horizontal), the grid size is 
$1.82\times2.84 \mbox{ km}^2$. Calculations at the resolution usually applied in this type of simulations amount to only a few horizontal gridpoints for a downflowing plume. Due to the increased number of gridpoints in our high resolution domain the simulations show the development of vigorous secondary instabilities of both the plume's head and stem.   Below a depth of about   $1\mbox{ Mm}$ the plume produces patches of low density, temperature, pressure and high vorticity which may last for all of our simulation time, $\sim 10 \mbox{ minutes}$, and probably considerably longer; they may be ascribed to the $2\mbox{D}$ nature of the present calculations. Centrifugal forces acting in these patches counteract the strong inward pressure. Probably most importantly, the plume's instabilities give rise to acoustic pulses created  predominantly down to $\sim 1.5\mbox{ Mm}$. The pulses proceed laterally as well as upwards and are ubiquitous.  Ultimately most of them emerge into the photosphere.  A considerable part of the photospheric `turbulence' in these models is due to those pulses rather than to some sort of eddies. -- The upflows in granules  are smooth where they reach the photosphere from below even in the present calculations; however, the pulses may enter in the photosphere also in granular upflows.
\end{abstract}


\section{Introduction}

Considerable activity is currently going on aimed at observing the solar atmosphere in high resolution. This is true for ground based observations (e.g. the Swedish vacuum telescope in Teneriffe), balloon borne instrumentation as in the Sunrise project, see \citet{schue1}, and satellite observations (e.g. the recently started HINODE satellite). The observations performed or planned with such facilities will contribute to a better understanding of hydrodynamic and magnetohydrodynamic phenomena near the Sun's surface. It is mandatory that these observational efforts be paralleled by simulations of truly high resolution, having  in mind ultimately the same goal.

We want to make a step in this direction. Current simulations of solar granulation and the atmosphere above it typically seem to apply a (horizontal) grid spacing down to about  $25 \mbox{ km}$, e.g.  \citet{stei98} or \citet{steff07}. Naturally, 3D simulations prevail nowadays. Since with decreasing mesh size the computational load becomes quite soon prohibitive it is tempting to refine the computational grid in a localized region of interesting processes only, using a normal grid spacing elsewhere. The technique of grid refinement is quite customary in other areas of astrophysical simulations, for example in supernova research as in \citet{plewa1}, but does not seem to have previously been applied in the solar context.

When it matters to achieve high resolution the numerical method should also be chosen with that goal in mind. We have adopted high-resolution schemes of the essentially non oscillatory (ENO) type.  An overview over ENO methodology can be found in \citet{harten1}. With high accuracy these methods can handle smooth flow, steep gradients and discontinuities at the same time. They can also be run largely without resorting to artificial diffusivities. 

Our specific approach has its pros and cons. The most basic contra is   the restriction of our simulations to two spatial dimensions. While this can and must be considered a drawback, there are, on the other hand, major advantages coming along with this decision. We can, in particular, easily attain a spatial resolution in the refined area which is out of reach in 3D; the present calculations adopt a grid size of $1.78\times2.84 \mbox{ km}^2$ (vertical $\times$ horizontal) in the high resolution area. Compare this grid spacing with the presently typical values mentioned above and the resolution of the solar optical telescope at the HINODE satellite: $0.25   \mbox{ arcseconds} \simeq 175\mbox{ km}$. (To be fair, we have to keep in mind in such a comparison   that a single computational grid point cannot represent a physical feature.) The 2D models can already give hints towards  phenomena which might occur in real solar granulation and can suggest ways to look at observations. They can aid in  more meaningful decisions what precisely should be investigated in 3D simulations. Note in addition that, even if 2D and 3D calculations naturally do not yield identical results, they still exhibit quite some similarity, see the discussion in \citet{aspl1}. As a consequence, even today 2D investigations are conducted with specific goals in mind, e.g. in \citet{wedem1}. Of course, the ultimate investigations of this sort will have to be done in $3\mbox{D}$; it is conceivable, however, that some of the findings are such that they may be encountered in $3\mbox{D}$ as well, in particular the results of section 3.2. 

Naturally, such calculations should in the long run additionally include magnetic fields. The present calculations consider the purely (radiation-) hydrodynamic case. They thus apply predominantly to the quiet Sun. It can be considered useful to perform both HD and MHD simulations in order to assess which phenomena must basically be ascribed to hydrodynamics and which ones to magnetohydrodynamics proper. 

In this paper, we direct our attention primarily to the downflowing plumes and the processes they excite. After all, these plumes are the most vigorous entities in solar granulation, have been encountered in any numerical model and are obviously a major ingredient to the overall dynamics. In addition, they are expected to be crucial in exciting solar p-mode oscillations, see \citet{rast99} and the references there. In many simulations addressing either  solar granulation or more idealized models of compressible convection, these plumes penetrate down several pressure scale-heights. Deeper down, by possibly joining with other plumes in a specific manner, they give rise to mesogranulation, \citet{stei98}, and perhaps even supergranulation. 

\section[]{Numerical Methods and physical model}

\subsection[]{Numerical Methods}

We have developed the package ANTARES (A Numerical Tool for Astrophysical RESearch) for simulations of solar granulation and other astrophysical flows. A more detailed description of ANTARES will be given elsewhere \citep{muth07}.

In the mode which is appropriate here ANTARES allows the numerical integration  of the continuity equation and the equations for momentum and energy balance in 2D (also in 1D or 3D) in time. The equations are solved in conservative form. The numerical scheme applied for the present calculations is the $5^{\mbox{th}}$ order accurate weighted ENO scheme of \citet{liu1}. In fact, we have various, quite different ENO schemes implemented and checked against standard examples and against each other; \citet{muth07}. -- Time integration is achieved using  specific second or third order Runge-Kutta schemes that preserve the properties of ENO schemes.

We solve the equation of radiative transfer by the method of short characteristics, see \citet{mihalas1}. This method makes ample use of interpolations which are done to at least second order of accuracy. At high optical depth we switch to the diffusion approximation, as appropriate. Values for opacity coefficients and the equation of state are taken from the OPAL tables, \citet{opal} augmented by line opacities based on the \citet{kurucz92} tables; the data as we need them are generated by an extension of the  \citet{piskunovkupka01} code. Wavelength variation of the opacities is accounted for by the method of opacity binning according to \citet{nord82}, \citet{ludwetal94}. This amounts to group into the same bin frequencies $\nu$ for which optical depth $\tau_ 
\nu=1$ is reached at similar geometrical depths.

Fixed, slip-free plates provide the boundary conditions at the top and at the bottom. At the bottom, the amount of energy flux appropriate for the solar case  enters the domain via radiative diffusion. 

The fixed boundary conditions at the top lead, in some instances, to a reflection of pulses. Still, since they are applied high in the atmosphere at low densities, these artificial effects are felt only in the higher atmosphere and not in the bulk of our domain. No major unwanted influence is exerted by the lower boundary condition.

\subsection[]{Physical and numerical parameters}

The basic computational domain without grid refinement encomprises a size of $2.78\times 11.19 \mbox{ Mm}^2$ (depth$\times$width) at a resolution of $382\times985$ equidistant grid points. The dimensions of a numerical cell  consequently are  $7.28\times11.36\mbox{ km}^2$. The density at the upper boundary of the whole domain is $\sim 2\times10^{-9}\mbox{g cm}^{-1}$. This corresponds to a location with a small optical depth in the continuum ($\sim 10^{-4}$) according to any standard solar model atmosphere.  The basic model was relaxed to a flux constancy of $\sim 1\% $ and then the high-resolution part was plugged in.

The refined region measures $1.97\times2.58\mbox{ Mm}^2$ with $1081\times909$ gridpoints, yielding a size of  $1.82\times2.84\mbox{ km}^2$ for the numerical cell. The high resolution domain starts also quite high in the atmosphere and spans $\sim 7$ pressure scale heights downwards.

Radiative transfer makes use of 12 bins. We switch smoothly from full radiative transfer to the diffusion approximation  at an optical depth of $\tau_{Ross}\sim 10000$.

Regarding the diffusion of momentum we have performed two sets of calculations. In model 2  a very small diffusion term  is included  in the momentum  equation (only) such that for the Prandtl number $\mbox{Pr}$ we have formally $\mbox{Pr}<0.01$ everywhere. In model 1 we have added diffusion terms in the momentum and energy equations according to the artificial diffusion description as in \citet{voe1}, who again refer to \citet{stei98}. This procedure amounts to adding a hyperviscous term and a term aimed at broadening of shocks to the equations.  This leads to an effective Prandtl number still quite small in the upper regions of the simulated domain; however, for model 1 the effective Prandtl number gets $O(1)$ in the lower two thirds of our computational box, in particular at places of strong velocity gradients. -- On a quantitative level, these two models differ to some extent. They agree, however, on the more qualitative level which is relevant for the present paper. Since model 2 has not been run quite as long as model 1 the presentation given here focusses on model 1.

\section{Discussion of the simulations}

\subsection{The plumes}

\subsubsection{The plume head}

The generation of new solar granules often commences with the appearance of a more or less circular spot of low temperature and high density forming  through cooling effects in the middle of a preexisting granule. Due to the rapid decrease of opacity with decreasing temperature this cooling constitutes a runaway process and initiates a downflowing plume of dense, cool material. In the real Sun, i.e. in $3\mbox{D}$, the plume later on frequently extends  sidewards along lanes and causes in this way a disruption of the original granule. The plume tends to be surrounded by more gentle upflows.

Considering now the evolution of a plume according to our calculations, we start with a discussion of the plume's head. In Fig.~1, three main plumes are visible. Somewhat to the left of the center there is a nascent plume. The nascent plume's  boundary and, by the way, also the lower boundary of the photosphere above granular upflow are exceedingly smooth and sharp even at this resolution, judging from specific entropy, which is plotted in the figure, and also from other relevant quantities. This finding is in agreement with $3\mbox{D}$ results obtained previously at common resolution as discussed above \citep{stei98}.-- The next plume to the left is in an advanced stage of stem instability (see the next subsection). Still farther left again is a plume in an earlier stage of development, exhibiting head instability. This instability comes about in the following way. Very soon after plume formation there is a tendency for the head to be faster than the subsequent material (and to broaden at the same time). Sidewards of the following tail low-density regions may easily develop. Their material flows in part towards the rear side of the head and gets entrained by the head's sideward lobes. In the course of time these rapidly spinning regions of low density detach from the plume proper and remain in appreciable depth ($>1\mbox{ Mm}$) for quite some time, about $10\mbox{ minutes}$ as much as we can tell and possibly considerably longer, although the contrasts to the surrounding material are really large, since its density may be only half of the outside value. These patches of high vorticity and low density resist pressure via Coriolis forces. They  show no tendency to ascend and to appear in the photosphere within a granular upflow. For further discussion of these vortex patches see subsection 3.1.2.

Instabilities of downward plunging plumes in compressible, layered media have been analyzed by \citet{rast1}. The settings in that paper were idealized, the initial structure of the layer being related to polytropes. In particular, \citet{rast1} investigated the case of low and high Prandtl number $\mbox{Pr}$. For  $\mbox{Pr}$ small ($\mbox{Pr}\leqq 0.1$) the head is a small, essentially circular, rotating structure which eventually detaches from the tail, with a subsequent similar structure appearing again and detaching from what is now the plume's head. On the other hand, the head's structure in the high-Prandtl-number case is much more similar to what we observe. Since we have a low Prandtl number case (using only the realistic radiative conductivity and, in some calculations, no artificial diffusivity) the question about the cause of that discrepancy arises. 

For a resolution of that issue consider the following. Firstly, as judging from figure $3$ of \citet{rast1}, the horizontal extent of the initial disturbance assumed there seems to be quite small compared to the pressure scale height. In our case, the cool spot which develops initially in the original granule is quite broad, of the order of a pressure scale height; in addition, its horizontal temperature profile does not resemble a peaked Gaussian but is trough-like instead. Furthermore, we have a general velocity field present, which will, via shearing forces, efficiently act on this broad initial disturbance, facilitating the development of further instabilities. We conclude that for these reasons we do not necessarily have to expect agreement with the results of \citet{rast1}.

\subsubsection{The plume stem}

Across the plume stem and in particular at the boundary to the surrounding material strong velocity gradients are present. As a consequence, as soon as the plume stem has attained some length a strong Kelvin-Helmholtz instability sets in. If surrounding disturbances are not overly large, the plume has a chance to initially develop a sinusoidal, lateral disturbance. Very soon, a von Karman vortex street appears which is well known to be a typical $2\mbox{D}$ instability. These vortices (as well as similar vortices which have been formed in the case of more severely disturbed plumes) are rapidly spinning, nearly circular patches of low density gas. As the plume disrupts due to increasing instability the vortices get separate entities, quite similar to the vortex patches which have been created out of the head's instabilities.

Such vortex patches are already visible in our large, unrefined domain. Once the motions in our refined subdomain have become sufficiently turbulent, several of them are seen there at each instant of time. The truly persistent vortex patches are located in the lower half of our domain. 

In the context of idealized microphysics, similar phenomena have been observed by \citet{port94} in their simulation of $2\mbox{D}$ convection. These authors assume, in particular, a gamma-law gas and a constant heat conduction coefficient. We agree in the solar context also on their observation, according to which density variation from the boundary to the center of a vortex patch is much larger than temperature variation. In a specific case we have found, for example, a density contrast of about a factor $2$ between patch center and the ambient medium, whereas temperature changes  by no more than $\sim 20 \%$. 

It should be noted, however, that those vortex patches are likely to be specific for the $2\mbox{D}$ case. They are not known in the $3\mbox{D}$ case. In particular, they do not seem to show up in our $3\mbox{D}$ high resolution models at least as far as these have been evolved up to now (i.e., not yet for a long physical time). In our $3\mbox{D}$ calculations we apply a grid refinement strategy similar to that one of the present paper and achieve, in the high resolution subdomain, a grid-size of $7.59\times\ 9.80\times 9.80 \mbox{ km}^3$ ($7.59\mbox{ km}=\mbox{vertical})$; see \citet{muth07}.

As a consequence of all those instabilities, the undisturbed  plume proper does not reach very deep (i.e., it usually ends, as far as we have observed, within our high-resolution subdomain). Still, plumes may nevertheless be embedded in broader, more gentle downflows, which reach essentially down to the lower boundary of our whole computational domain and would extend presumably even deeper if our domain would only extend farther down. Therefore, the increased instabilities of the plumes which we see in high resolution do not  contradict the findings by  \citet{stei98} that downflows merge at large depth, forming the mesogranulation and possibly even the supergranulation network.

As long as the plume's stem fairly maintains its integrity the velocities of the plume's core as measured in the fixed coordinate system tend to exceed the local velocity of sound. 
\subsection{Acoustic pulses and atmospheric turbulence}

The downflowing plumes generate acoustic pulses. To a considerable part, pulse generation is associated with the instabilities of the plume's head and stem. For example, a stem may experience a strong lateral displacement, say to the right.  Then  the rapidly downflowing material can happen to collide with material belonging to a gentle upstream to the right of the stem thus causing a density, temperature and pressure enhancement. The location  of  disturbance generation will, in general, move during the course of time. As the disturbance signals are continuously created at that moving point and propagate with sound speed they appear initially essentially as a  straight or curved line, depending on sound speed variations and possibly also variations in the velocity of the source or the ambient medium. 

The initial size (length of a front) of a pulse is largely set by the source's lifetime. As a consequence, pulses generated in the complex inflow field near the photosphere (where the pulses, in addition, easily escape upwards) may be localized; also, in the late stage of plume development where the integrity of the plume has been lost small or not so regular pulses may be created. As is evident from the generation mechanism described above these pulses are formed where flow fields of the necessary characteristics are present, i.e. mainly from somewhat below the photosphere down to a depth of $\sim1.5\mbox{ Mm}$. Below about that depth the plumes are, in general, no longer sufficiently vigorous in order to generate pulses.

Following the pulse front is a region of relatively high density and temperature, at least near its origin. Density and temperature are next to discontinuous across the pulse. The pulses can move against the local flow field and are therefore not advection phenomena but akin to sound waves.

Due to strengthening mechanisms (see below) some pulses can travel over surprisigly long distances, some of them at least a few granule widths. -- Pulses  are ubiquitous.  At each instance of time there is a considerable number of them in our high-resolution computational domain, in particular also in the atmosphere, see Fig.~2, once the solution has only been given time to evolve away from the more laminar initial state taken from the coarse grid. Through interaction with the complex inflow field of a granule the plumes may get enhanced just there so that chances are good to see very strong pulses (quite possibly more  than one at a time) moving from below into the photosphere above a downflow, contributing to a very considerable part to turbulence (gradients of velocity etc.) there. They run, however, into the photosphere also in otherwise smooth granular upflows.

The pulses, travelling with sound speed, exhibit in part linear, in part nonlinear character. Two oppositely moving pulses running into each other penetrate and continue their course more or less undisturbed, in the sense of wave superposition, which can also occur for wave packets. The same may also be true to some degree if a pulse traverses the more regular, not chaotic, part of a plume, although we believe to have seen plume strengthening as a consequence. 

Predominantly when a pulse hits a part of a plume with a complicated flow field, for instance in the lower part of a developed plume or in the atmospheric inflow region at the top, and interacts with the complex pattern of  motions prevailing there, the interaction is nonlinear. In addition to the obvious distortion due to the variation in sound speed across a plume together with the effects of the plume's shape and flow field, it may trigger the generation of subsequent pulses or may be strengthened.  

Because of strengthening it may cross several downflowing plumes and  maintain its identity over several granule widths and traverse, e.g., all the width of our high resolution domain, $>2 \mbox{ Mm}$. Indeed, the dominant pulse in Fig.~2 (the semicircle in the left upper part of the figure) is an example for the nonlocality of pulses. It has entered the high-resolution domain from the right as a quite feeble disturbance and gained in strength through repeated interaction with the complicated flow field of the plumes it crossed. Other pulses may be more local, among others because they have been directed mainly upwards right from the beginning.
 
On the other hand, pulses do not seem to get easily absorbed when crossing plumes. We have seen, however, a reduction of the number of pulses procedding in the following way. A leading pulse leaves behind it a region of higher sound speed. When a trailing pulse is about to overtake the leading one these two pulses merge into one. --  The pulses are noticeably refracted when entering the photosphere from below at some angle because of the sharp drop in sound velocity.

The pulses and their generation mechanism obviously are not identical to the acoustic events discussed in \citet{skart00}. They found pulses which are generated when a small granule collapses and a larger downflow is initiated  at the site of the collapse. Given the relatively rare starting configuration one may presume that these pulses are considerably less frequent than the acoustic pulses discussed here. -- According to our evidence the collision of plumes, which seems occasionally to have been held responsible for the generation of pulses, cannot be considered a major source of pulses at least from the viewpoint of frequency of events.

These phenomena (secondary instabilities of the plumes, pulse generation) are most clearly seen in the high-resolution part of our calculation. It is remarkable that, however, they are also distinctly present in the low resolution portion. We take this as an indication of the quality of the basic numerical scheme.

It will be of interest to look into the material in a more quantitative fashion in order to figure out what these instabilities respectively pulses imply for the transport properties of convection and for questions of heating of the chromosphere  in the quiet sun. Similarly, the possible role of the acoustic pulses in the excitation of p-modes needs to be investigated. Of course, due to the main limitation of the present models, namely that they are $2\mbox{D}$, it will be necessary to perform analogous simulations for the  full $3\mbox{D}$ case. As mentioned previously, $3\mbox{D}$ calculations are presently being pursued. They apply a grid spacing of $7.59\times\ 9.80\times 9.80 \mbox{ km}^3$ in the high resolution region, thus similar to the low resolution portion of the present $2\mbox{D}$ simulations, and will be reported in a forthcoming paper.

\section*{Acknowledgments}

This research was supported by the Austrian Science Foundation, project AP17024. We thank the referee for comments which helped to clarify the presentation.

\begin{figure*}
\begin{minipage}{130mm}
\begin{center}
\scalebox{1}[1]{\includegraphics[width=125mm]{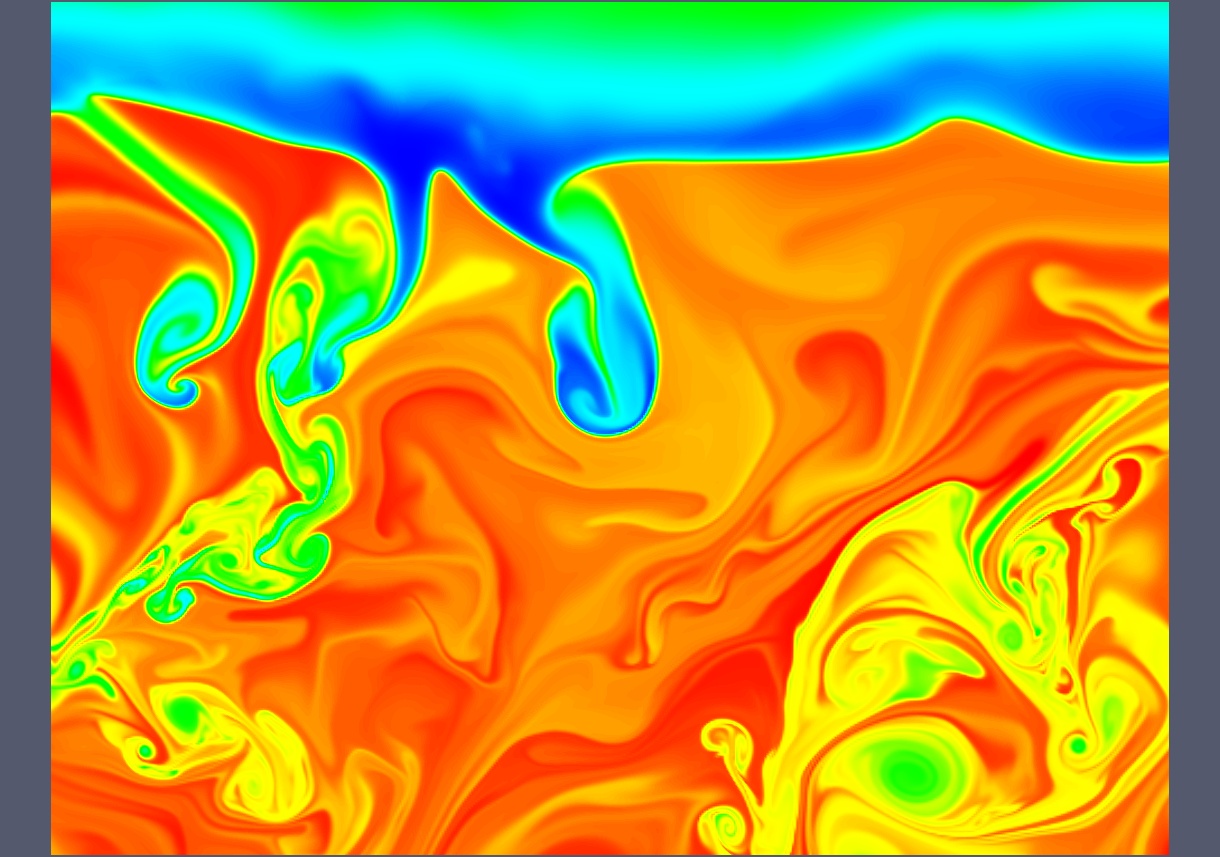}} 
  \label{fig1}
	\caption{Specific entropy (red = high) in the high-resolution domain. Note the sharp and smooth boundary of the nascent plume (left of the middle); the instabilities and entrainment of surface material (low entropy) in the more developed plume (farther left); and the head instabilites of the plume quite to the left.  Compare with the smooth structures in the entropy plot in 
	\citet{stei00}, fig. 11 there. In vorticity indications of  turbulence are, however, seen there, l.c., fig. 13.}
\end{center}
\end{minipage}
\end{figure*}

\begin{figure*}
\begin{minipage}{130mm}
\begin{center}
 \scalebox{1}[1]{\includegraphics[width=125mm]{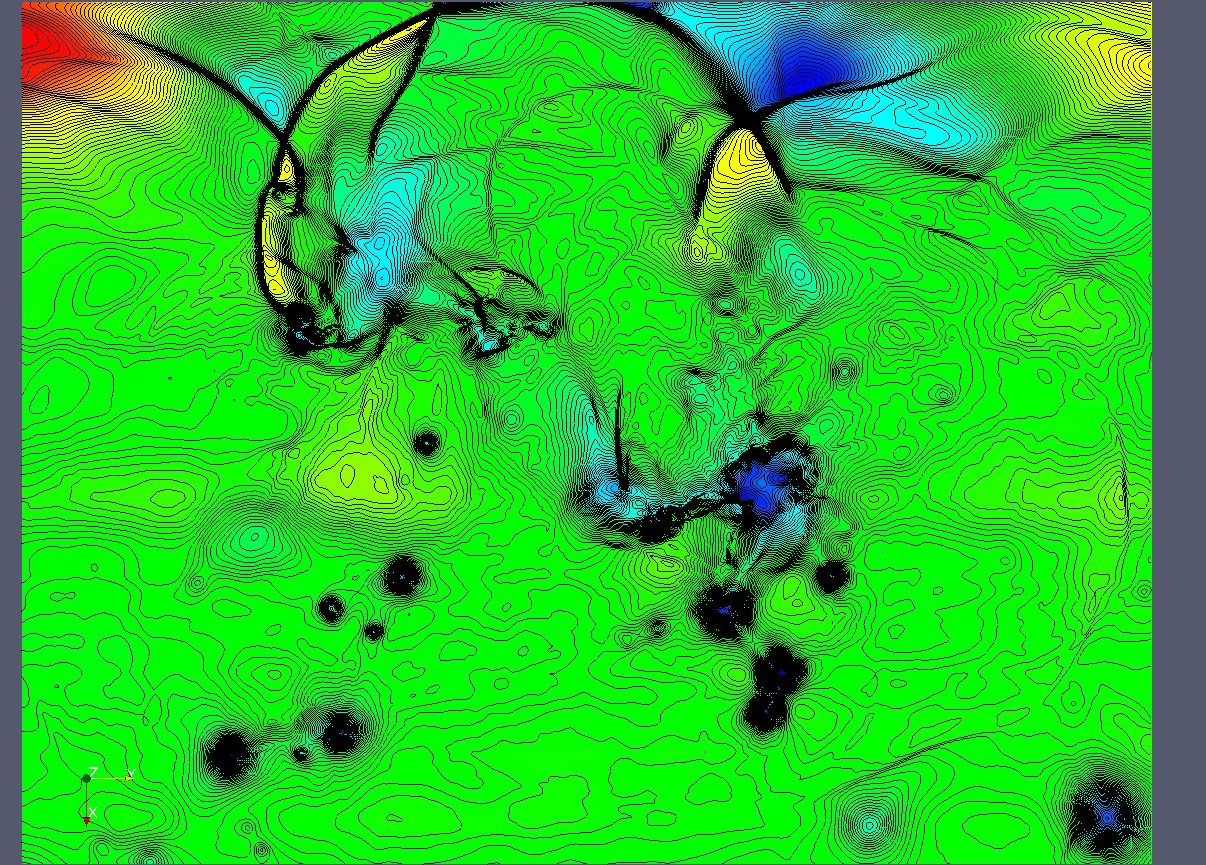}} 
  \label{fig2}
	\caption{Snapshot of the high resolution region with the plumes in a more developed state. Color: $\log{(\rho/\bar{\rho})}$ ($\rho = $ density, $\bar{\rho}=$ horizontal average of density; red = high density. Isolines of $\log{(p/\bar{p})}$ (black), where $p =$ pressure,  crowd near acoustic pulses and shocks. The left half of the large pulse (semicircle) moves to the left, the other pulses crossing it there move to the right. A number of circular patches of low density and pressure and high vorticity can be discerned in the lower half of the figure.}
\end{center}
\end{minipage}
\end{figure*}

\label{lastpage}

\end{document}